\documentclass[aps,pra,twocolumn,showpacs]{revtex4}
\usepackage{graphicx}
\usepackage{amsmath}

\usepackage{bbm}

\newcommand{\abs}[1]{\left|{#1}\right|}
\newcommand{\br}[1]{\langle #1|}
\newcommand{\ke}[1]{|#1\rangle}

\newcommand{\da}{^\dagger}
\newcommand{\pt}[1]{\left( #1 \right)}
\newcommand{\pq}[1]{\left[ #1 \right]}
\newcommand{\pg}[1]{\left\{ #1 \right\}}

\newcommand{\al}[1]{^{(#1)}}

\begin{document}

\title{Resonance fluorescence of a cold atom in a high-finesse
resonator} \author{Marc Bienert,$^{1}$ J. Mauricio Torres,$^{1}$
Stefano Zippilli,$^{2,3}$ and Giovanna Morigi$^2$} \affiliation{
$^1$Instituto de Ciencias F{\'i}sicas, Universidad Nacional Aut{\'o}noma de M{\'e}xico, 62251 Cuernavaca, Morelos, M{\'e}xico\\
$^2$ Grup d'Optica, Departament de Fisica, Universitat Aut\`onoma de Barcelona, 08193 Bellaterra, Spain\\
$^3$ ICFO - Institut de Ci\`encies Fot\`oniques, 08860 Castelldefels (Barcelona), Spain
}

\date{\today}  \begin{abstract} We study the spectra of emission
of a system composed by an atom, tightly confined inside a
high-finesse resonator, when the atom is driven by a laser and is
at steady state of the cooling dynamics induced by laser and
cavity field. In general, the spectrum of resonance fluorescence
and the spectrum at the cavity output contain complementary
information about the dynamics undergone by the system. In certain
parameter regimes, quantum interference effects between the
scattering processes induced by cavity and laser field lead to the
selective suppression of features of the resonance fluorescence
spectrum, which are otherwise visible in the spectrum of
laser-cooled atoms in free space.

\end{abstract} \pacs{32.80.Pj, 42.50.Vk} \maketitle

\section{Introduction}

The modification of atomic dynamics by means of optical resonators
is actively investigated by several groups worldwide. Several
studies have recently focused on the manipulation and control of
the atomic center-of-mass motion by mechanical forces induced by
the resonator. Recent experiments demonstrated mechanical forces
of single photons on single atoms
\cite{Pinkse00,Hood00,Bushev03,Bushev04}, cooperative effects and
selforganization of laser cooled atomic clouds in
resonators~\cite{Chan03,Zimmermann04,Hemmerich03}, and
cavity-assisted cooling of atoms~\cite{Pinkse04,KimbleRaman}. The
atomic motion and resonator field states are interdependent, as
the atomic motion is determined by the mechanical forces of the
resonator's field, while the resonator's field is strongly
affected by the medium's refractive index and thus by the atom's
position and velocity~\cite{RitschDomokosJOSA}. Information about
the atomic motion can be obtained through the emitted
light~\cite{Bushev04,OpticalKaleydoscope}, and has been used to
implement feedback-induced trapping and cooling of the atomic
motion~\cite{Fisher,Bushev06}. In general, the quantum behaviour
of atoms in resonators is still largely unexplored.

In this manuscript we study theoretically the spectrum of
resonance fluorescence and the spectrum at the cavity output of
the light scattered by an atom, tightly confined inside an optical
resonator and continuously driven by a laser. The photons
collected at the detectors are scattered by the atom at the steady
state of cavity-assisted cooling, and carry the information of the
coupled dynamics between cavity and atomic external quantum
degrees of freedom. We analyze the spectra in various cavity
cooling regimes, which have been identified
in~\cite{Vuletic,Zippilli05}, applying and extending the method
in~\cite{Bienert04,Bienert06}, used for evaluating the spectrum of
resonance fluorescence of a laser cooled atom, to our case, in
which the atom is strongly coupled to a mode of the quantum
electromagnetic field. This work extends and complements previous
theoretical investigations on the property of the light scattered
by atoms inside resonators, see for
instance~\cite{Mondragon,Carmichael85,Bonifacio,Walls,CarmichaelSavage89,Carmichael92,Kimble94,Quang,Rice04},
by considering systematically the coupled dynamics between field and quantized atomic motion, 
which determines the steady state and the
properties of the scattered light. Differing from studies on the
coupled dynamics of atoms in resonator, where both confinement and
cooling are provided by the cavity field
potential~\cite{Domokos02,Salzburger04}, in this work the atom is
confined by an external potential, like an ion or a dipole trap,
and its motion is cooled by the mechanical effects of laser and
cavity mode field.

This work is organized as follows. In Sec.~\ref{Sec:Spectra} the
setup of the system is discussed. In Sec.~\ref{Sec:Model} the
theoretical model is introduced and in Sec.~\ref{Sec:Results} the
spectra of emission are reported and discussed. Finally, in
Sec.~\ref{Sec:Conclusions} the conclusions and outlooks are
presented.

\section{Spectra of emission} \label{Sec:Spectra}

\begin{figure}[!th] \includegraphics[width=6cm]{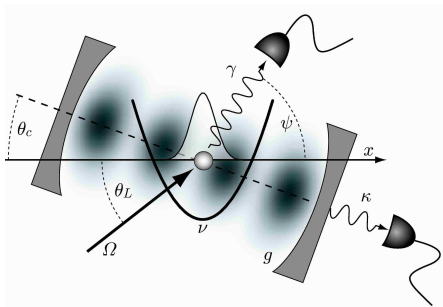}
\caption{An atom is confined by an external potential inside an
optical resonator. The motion is essentially along the $\hat{x}$
direction (the radial potential is assumed to be very steep) and
is harmonic at frequency $\nu$. A mode of the resonator couples
with strength $g$  to the dipole, which is driven transversally by
a laser at Rabi frequency $\Omega$. The system dissipates by
spontaneous emission of the atomic excited state at rate $\gamma$
and by cavity decay at rate $\kappa$. Two detectors measure
respectively the field at the cavity output and the atomic
fluorescence at an angle $\psi$ with respect to the axis of the
motion. $\theta_c$ and $\theta_L$ indicate the angles of the
cavity and laser wave vectors with the motional axis. }
\label{Fig:1} \end{figure}

We consider an atom, whose dipole transition couples to the mode
of a high-finesse resonator and is driven by a laser, in the setup
sketched in Fig.~\ref{Fig:1}. The atomic motion is confined by a
tight harmonic potential and is kept cooled by the mechanical
effects of cavity and laser fields. Two detectors record the
photons scattered by the system, which is at the stationary state
of the cooling process: One detector reveals the spectrum of the
intensity of the spontaneously scattered photons, another detector
measures the power spectrum of the photons at the cavity output.

The corresponding signals are, apart from a constant
proportionality factor, the Fourier transform of the
auto-correlation function of the electric field, and take the form
\begin{eqnarray} {\bf S}\al{j}(\omega)\propto{\rm
Re}\int_0^{\infty}{\rm d}\tau{\rm e}^{-{\rm i}\omega\tau} \langle
E_j^{(+)}(t+\tau)E_j^{(-)}(t)\rangle \end{eqnarray} where
$E^{(+)}(t)$ and $E^{(-)}(t)$ are the negative and positive
frequency component of the electric field at the detector $j$ at
time $t$, and $\langle\cdot\rangle$ describes the average over the
electromagnetic field, the atomic dipole and center-of-mass
degrees of freedom. From now on we label with the superscript
"$j=$at" the signal of resonance fluorescence and with "$j=$cav"
the signal at the cavity output. In the vacuum the intensity of
the scattered field is proportional to the source field. Hence, in
the far-field of the dipole source the electric field is
proportional to the retarded value of the atomic dipole, $E_{\rm
at}^{(-)}(t)\propto D(t-t')$, with $D$ dipole operator, and at the
cavity output is proportional to the cavity field, $E_{\rm
cav}^{(-)}(t)\propto a(t-t')$, with $a$ annihilation operator of a
photon of the cavity mode. The power spectra at steady state are
defined as \begin{eqnarray} \label{SS} {\bf
S}^{(j)}(\omega)=\chi\al{j}{\cal S}^{(j)}(\omega) \end{eqnarray}
where $\chi\al{\rm at}$ and $\chi\al{\rm cav}$ are prefactors
which scale with the spontaneous decay rate $\gamma$ and the
cavity decay $\kappa$, respectively. In the reference frame of the
laser at frequency $\omega_L$, the spectral form is given by
\begin{eqnarray}  &&{\cal S}^{\rm (at)}(\omega)={\rm
Re}\int_0^{\infty}{\rm d}\tau{\rm e}^{-{\rm
i}(\omega-\omega_{L})\tau} \langle D^\dagger(\tau)
D(0)\rangle_{\rm st},  \label{Sat}\\ && {\cal S}^{\rm
(cav)}(\omega)= {\rm Re}\int_0^{\infty}{\rm d}\tau{\rm e}^{-{\rm
i}(\omega-\omega_{L})\tau} \langle a^\dagger(\tau)
a(0)\rangle_{\rm st}, \label{Scav} \end{eqnarray} where the
average $\langle \cdot\rangle$ is evaluated by tracing over the
atomic and cavity degrees of freedom, which are treated quantum
mechanically. In this work we systematically consider the coupling
of the cavity field with the quantized center-of-mass motion in
determining the cooling dynamics and steady state, and its
manifestations in the transmission spectrum and in the spectrum of
resonance fluorescence. In particular, we focus on the peculiar
properties which are due to the presence of the resonator. At this
purpose, we shall give a brief review of some basic properties of
the spectrum of a trapped and laser-cooled atomic dipole in free
space.

The processes leading to cooling of an atom confined by a harmonic
trap and in free space give rise to sidebands of the peaks of the
spectrum of resonance fluorescence. They result from the
mechanical processes associated with the change of the motional
state by one excitation. The most evident features are the
sidebands of the elastic peak, which contain most information on
the cooling process \cite{Cirac93,Lindberg86,Raab00}. In
particular, for a two-level atom cooled by a laser at frequency
$\omega_L$, the lower sideband of the elastic peak, at frequency
$\omega_L-\nu$, corresponds to photons which heat the motion of
one excitation, while the upper sideband, at $\omega_L+\nu$, to
photons which cool the motion. Cooling is achieved by enhancing
the rate of scattering of cooling photons over the heating
photons. At steady state the number of heating and cooling photons
per unit time is equal and, below saturation, the two sideband
signals integrated over all solid angle of emission are Lorentzian
curves of same height and width~\cite{Lindberg86}. Nevertheless,
at a specific detection (emission) angle, the sidebands of the
elastic peak are asymmetric in height and in their functional
dependence on the frequency. This behaviour is due to quantum
interference between paths of excitations coupling internal and
external degrees of freedom~\cite{Cirac93}. In~\cite{Bienert06} it
has been shown that a similar behaviour can be observed for the
sidebands of the inelastic spectrum. \begin{figure}[!th]
\includegraphics[width=5cm]{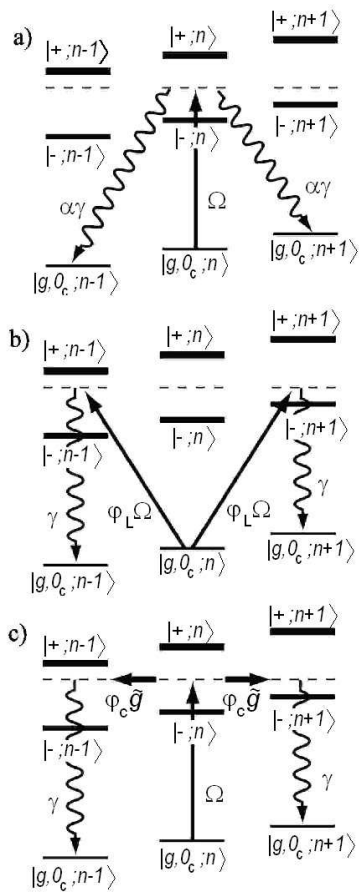} \caption{Scattering
processes, where the photon is emitted by spontaneous emission,
leading to a change of the vibrational number. The states
$|g,0_c;n\rangle$, $\ke{\pm;n}$ are the cavity-atom dressed states
at phonon number $n$. The parameters $\alpha,\varphi_c,\varphi_L$
emerge from the mechanical effects of light and are defined in
Sec.~\ref{basicEq}.} \label{Fig:3} \end{figure}
\begin{figure}[!th] \includegraphics[width=5cm]{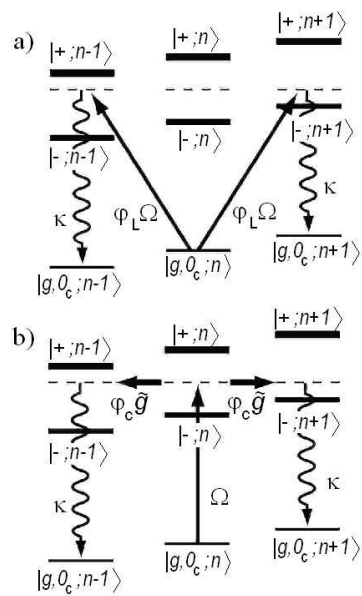}
\caption{Scattering processes, where the photon is emitted by
cavity decay, leading to a change of the vibrational number.}
\label{Fig:3b} \end{figure}

A high-finesse resonator modifies significantly the scattering
properties of the atom. In the regime of strong coupling the
resonator saturates the dipole transition at the single photon
level~\cite{Kimble94}. This property gives rise to inelastic
features of the resonance fluorescence spectrum, which are already
visible when the dipole is driven by a weak laser~\cite{Quang}.
When additionally the atomic motion is considered, the presence of
an optical resonator can give rise to a rich variety of
interference effects, which, among other, affect the features of
the motional sidebands. In the limit of weak drives the relevant
scattering processes which cool and heat the atomic motion are
depicted in Figs.~\ref{Fig:3} and \ref{Fig:3b}~\cite{Zippilli05}.
In the figures, the levels coupled by the laser are the stable
state $|g,0_c,n\rangle$ and the dressed states $\ke{\pm,n}$, which
are the eigenstates of dipole-cavity interaction and are
superpositions of the states $\ke{g,1_c,n}$ and $\ke{e,0_c,n}$.
Here, $|g\rangle$ and $\ke{e}$ are the electronic ground and
excited states of the dipole transition, $|0_c\rangle$ and
$\ke{1_c}$ are the cavity mode state with 0 and 1 photons, and
$|n\rangle$ denotes the vibrational excitations.
Figure~\ref{Fig:3} describes the contributions which add up
coherently when the atom is scattered in a different motional
state, $n\to n\pm 1$, and the photon is dissipated by atomic
spontaneous emission. 
In Fig.~\ref{Fig:3}(a) the driving laser, at Rabi frequency $\Omega$, excites the system without modifying the motional state which, instead, is changed by the recoil induced by the spontaneously emitted photon. Here the mechanical effects are scaled by the geometric factor $\alpha$ which is found after averaging over the solid angle of spontaneous emission, see Sec.~\ref{basicEq}.  In Fig.~\ref{Fig:3}(b) the change in the center-of-mass state is due to the mechanical coupling with the laser and in Fig.~\ref{Fig:3}(c) with the cavity field at vacuum Rabi frequency $\tilde g$. The parameters $\varphi_L$ and $\varphi_c$, which are defined in Sec.~\ref{basicEq}, depend on the geometry of the set-up and scale the recoil effect of a laser photon and of a cavity photon respectively.
Fig.~\ref{Fig:3b} describes the corresponding processes when the
photon is dissipated by cavity decay. 
In~\cite{Zippilli05} quantum
interference between the processes depicted in Fig.~\ref{Fig:3}
has been used in order to reduce the ratio $A_+/A_-$ between the
rates of heating and of cooling, and therefore to enhance the
cooling efficiency. We remark that a transition
$|g,0_c,n\rangle\to |g,0_c,n\pm 1\rangle$ corresponds to the
emission of a photon at frequency $\omega_L\mp \nu$, and to a
corresponding signal in the spectrum of emission. Interference
between the scattering processes, leading to photon emission, will
give a corresponding modification of the sideband signals, which
will be generally different from the one observed for sideband
cooling in free space. We refer the reader to
Sec.~\ref{Sec:Results}, where the evaluated spectra are shown and
discussed.

\section{The Model}
\label{Sec:Model}

The spectra of emission in Eqs.~(\ref{Sat}) and~(\ref{Scav}) are
evaluated by averaging over the atomic and cavity degrees of
freedom, namely tracing over the density matrix $\varrho$ for the
cavity mode, dipole and quantized center-of-mass motion. The
two-time correlation function are determined by the Liouvillian
${\cal L}$, which gives the evolution of the density matrix,
$\dot\varrho={\cal L}\varrho$, and the averages are taken at the
stationary state of this evolution, using the density matrix
$\varrho_{\rm st}$ satisfying the equation ${\cal L}\varrho_{\rm
st}=0$. This corresponds to measure the spectra at the latest
stages of cooling, when cavity and atom are in the dynamical
equilibrium determined by the laser and the coupling to the
external environment.

The evaluation of the specific form of the spectra is made by
applying the spectral decomposition of the Liouville
operator~\cite{Bienert04,Bienert06,Briegel93,Barnett2000}, where
we use the solutions of the eigenvalue equations \begin{eqnarray}
{\cal L}\varrho^{\lambda}&=&\lambda\varrho^{\lambda},\nonumber\\
\check{\varrho}^{\lambda}{\cal L}&=&\lambda
\check{\varrho}^{\lambda}, \label{eq:seceq} \end{eqnarray} with
eigenvalues $\lambda$ and right and left eigenelements
$\varrho^{\lambda},\check{\varrho}^{\lambda}$, whose scalar
product is defined by the trace, and fulfill the orthogonality
relation ${\rm
Tr}\{\check{\varrho}^{\lambda'}\varrho^{\lambda}\}=\delta_{\lambda,\lambda'}$.
The Liouvillian is decomposed into the sum of projectors, ${\cal
L}=\sum_\lambda \lambda  {\cal P}^{\lambda}$, whose action on an
arbitrary operator $X$ is defined by ${\cal
P}^{\lambda}X=\varrho^{\lambda}{\rm
Tr}\{\check{\varrho}^{\lambda}X\}$. Accordingly, the spectra are
rewritten as \begin{equation} \label{Sat:1}{\cal S}\al{\rm
at}(\omega)={\rm Re} \sum_{\lambda}\frac{1}{{\rm
i}(\omega-\omega_{L})-\lambda} {\rm Tr}\left\{D^{\dagger} {\cal
P}^{\lambda}D\varrho_{\rm st}\right\} \end{equation} and
\begin{equation} \label{Scav:1} {\cal S}\al{\rm cav}(\omega)={\rm
Re} \sum_{\lambda}\frac{1}{{\rm i}(\omega-\omega_{L})-\lambda}
{\rm Tr}\left\{a^{\dagger} {\cal P}^{\lambda}a\varrho_{\rm
st}\right\}. \end{equation} In this section we outline the
procedure for evaluating these terms: in Sec.~\ref{basicEq} we
define the Liouvillian ${\cal L}$ for the system we study, in
Sec.~\ref{Sec:Steady} we briefly discuss the steady state of the
cooling dynamics, and in Sec.~\ref{spectra} we show how
Eqs.~(\ref{Sat:1}) and~(\ref{Scav:1}) are explicitly determined.

\subsection{Liouvillian}\label{basicEq}

The system we investigate consists of an atom with mass $M$ whose
center-of-mass motion is trapped by a harmonic potential of
frequency $\nu$. A dipole transition connecting the ground state
$\ke{g}$ to the excited state $\ke{e}$, at frequency $\omega_0$
and linewidth $\gamma$, is strongly coupled to a single mode of an
optical cavity at frequency $\omega_c$. The cavity field decays at
rate $\kappa$ due to the finite transmission of the mirrors. The
atomic transition is driven by a laser field at frequency
$\omega_L$, which, in combination with the resonator, cools the
motion. We restrict our description of the center-of-mass motion
to the $x$ axis and we indicate with $\theta_c$ and $\theta_L$ the
angles that the wave vectors, ${\bf k}_c$ and ${\bf k}_L$, of
cavity mode and laser form with the $x$ axis, see
Fig.~\ref{Fig:1}. The coupling between atomic dipole and
electromagnetic radiation is assumed to be in the Lamb-Dicke
regime, when the width of the atomic wave packet is small compared
with the field wavelength $2\pi/k$, where here $|{\bf
k_L}|\approx|{\bf k_c}|=k$. In this regime the mechanical effect
of light can be studied in perturbation theory in the Lamb-Dicke
parameter $\eta=k\sqrt{{\hbar}/{2M\nu}}$. At second order in
$\eta$ we express the Liouvillian, determining the evolution of
the density matrix of atom and resonator degrees of freedom, as
\begin{eqnarray}\label{meq} {\cal L}={\cal L}_0+{\cal L}_1+{\cal
L}_2+{\rm O}(\eta^3) \end{eqnarray} where ${\cal L}_j$ is the
corresponding order in the Lamb-Dicke expansion. At zeroth order
in $\eta$ the external dynamics is decoupled from the
dipole-cavity dynamics and \begin{eqnarray}\label{L0} {\cal
L}_0={\cal L}_{E}+{\cal L}_{I} \end{eqnarray} where
\begin{eqnarray} \label{LE} {\cal L}_E\varrho=-{\rm i}\nu\pq{b\da
b,\varrho} \end{eqnarray} is the Liouvillian for the
center-of-mass oscillator. Here $b$ and $b\da$ are the
annihilation and creation operators of a quantum of vibrational
energy. Term ${\cal L}_{I}$ in Eq.~(\ref{L0}) acts only on cavity
and dipole degrees of freedom and it is given by \begin{eqnarray}
{\cal L}_{I}\varrho&=& \frac{1}{{\rm i}\hbar}[H_{\rm at}+H_{\rm
cav}+H_{\rm at-cav}+H_{L},\varrho]+{\cal K}\varrho+{\cal
L}_s\varrho,\nonumber\\ \label{LI} \end{eqnarray} where
\begin{eqnarray}
&&H_{\rm at}=-\hbar\Delta\sigma\da\sigma\\
&&H_{\rm cav}=-\hbar\delta_ca\da a \end{eqnarray} with
$\sigma=\ke{g}\br{e}$, and $\sigma\da$ its adjoint, $a$, $a\da$
are the annihilation and creation operators of a cavity photon,
$\Delta=\omega_L-\omega_0$ and $\delta_c=\omega_L-\omega_c$ are
the detunings of the laser from the dipole and from the cavity
frequency, respectively. The terms
\begin{eqnarray}\label{H0at-cav}
&&H_{{\rm at-cav}}=\hbar \tilde{g}(a\da\sigma+a\sigma\da)\\
&&H_{L}=\hbar\Omega(\sigma\da+\sigma)\label{H0L} \end{eqnarray}
describe the radiative couplings of the dipole with the cavity
mode, at vacuum Rabi frequency $\tilde{g}$, and with the laser, at
Rabi frequency $\Omega$. The coupling constant
$\tilde{g}={g}\cos\phi$ contains the phase $\phi$ determining the
position of the trap center in the mode spatial function. The
incoherent dynamics at zeroth order in $\eta$ is described by the
two superoperators ${\cal K}$ and ${\cal L}_s$ in Eq.~(\ref{meq})
which account for the cavity decay and  dipole spontaneous
emission, \begin{eqnarray}\label{atomdecay} &&{\cal
K}\varrho=\frac{\kappa}{2}(2a\varrho a\da-a\da a\varrho-\varrho
a\da a)\\
&&{\cal L}_s\varrho=\frac{\gamma}{2}\left(
2\sigma{\varrho}\sigma\da
-\sigma\da\sigma\varrho-\varrho\sigma\da\sigma\right).
\end{eqnarray} The coupling between external and internal atomic
degrees of freedom enters in the term ${\cal L}_1$ and ${\cal
L}_2$ of Eq.~(\ref{meq}). The first order Liouvillian is
\begin{eqnarray} {\cal L}_1\varrho=-{\rm
i}\eta\pq{(b\da+b)V_1,\varrho} \end{eqnarray} with
\begin{eqnarray}\label{V} V_1=\varphi_c V_{1 c}+\varphi_L V_{1 L},
\end{eqnarray} where  \begin{eqnarray}\label{1stlaser}
V_{1L}&=&{\rm i}\Omega(\sigma\da-\sigma)\\
V_{1c}&=&-\tilde g(a\sigma\da+a\da \sigma) \label{1stcavity}
\end{eqnarray} describe respectively the interaction of the atomic
dipole with the laser and with the cavity field at first order in
the Lamb-Dicke parameter. The mechanical effects of the laser and
of the cavity field are scaled by the geometrical factors
\begin{eqnarray}
\varphi_L&=&\cos\theta_L\\
\varphi_c&=&\cos\theta_c\tan\phi.
\end{eqnarray}
The Liouvillian at second order is
\begin{eqnarray}
{\cal L}_2\varrho=-{\rm i}\eta^2\pq{(b\da+b)^2 (\varphi_c^2 V_{2c}+\varphi_L^2 V_{2L}),\varrho}+{\cal L}_{2s}
\end{eqnarray}
where
\begin{eqnarray}\label{V2}
V_{2L}&=&-\frac{\Omega}{2}(\sigma\da+\sigma)\\
V_{2c}&=&-\frac{\tilde g}{2 \tan^2\phi}(a\sigma\da+a\da \sigma).
\end{eqnarray} Superoperator ${\cal L}_{2s}$ accounts for the
recoil effect of photons scattered in the modes of the
electromagnetic field external to the cavity, \begin{eqnarray}
{\cal L}_{2s}\varrho&=\eta^2&\frac{\gamma\alpha}{2}\sigma\left[2(b+b\da)\varrho(b+b\da)\right.\nonumber\\
&&\left.-(b+b\da)^2\varrho-\varrho(b+b\da)^2\right]\sigma\da
\end{eqnarray} and $\alpha=\int_{-1}^1 {\rm d}\cos\theta
\cos^2\theta {\cal N}(\cos\theta)$ gives the angular dispersion of
the atom momentum due to the spontaneous emission of photons.
Finally, the dipole $D$ entering in Eq.~(\ref{Sat}) is decomposed
according to $D=D_0+D_1+D_2$, where
\begin{eqnarray*} &&D_0=\sigma,\\
&& D_1=-{\rm  i}\eta\sigma(b\da+b)\cos\psi,\\
&&D_2=-\frac{1}{2}\eta^2\sigma(b\da+b)^2\cos^2\psi
\end{eqnarray*}
and $\psi$ is the angle of the emitted photon with the motional axis.

\subsection{Steady state of the cooling dynamics}
\label{Sec:Steady}

The averages in Eqs.~(\ref{Sat}) and~(\ref{Scav}) are taken over
the stationary state of the system. This is described by the
density matrix $\varrho_{\rm st}$ satisfying ${\cal L}\varrho_{\rm
st}=0$, which is an eigenvalue equation for the operator ${\cal
L}$ at eigenvalue $\lambda=0$. Thus, $\varrho_{\rm st}$ is the
right eigenelement of ${\cal L}$ at $\lambda=0$, $\varrho_{\rm
st}\equiv \varrho^0$. At zero order in the Lamb-Dicke expansion
$\varrho_{\rm st}$ is the product of the density matrices for the
internal and external degrees of freedom, $$\varrho_{\rm
st}=\rho_{\rm st}\mu,$$ where $\rho_{\rm st}$ satisfy ${\cal
L}_{I}\rho_{\rm st}=0$, see Eq.~(\ref{LI}), and $\mu$ is the
density matrix of the external degrees of freedom, steady-state
solution of the cooling equation~\cite{Zippilli05}
\begin{eqnarray}
\dot\mu&=&\frac{\eta^2}{2}A_-\left[2b\mu b^{\dagger}-b^{\dagger}b\mu-\mu b^{\dagger}b \right]\nonumber\\
&&+\frac{\eta^2}{2}A_+\left[2b^{\dagger}\mu b-bb^{\dagger}\mu-\mu
bb^{\dagger} \right],\label{RateEq:pn} \end{eqnarray} with
$A_+<A_-$. In particular, \begin{equation} \label{mu}
\mu=\frac{1}{1+\langle
    n\rangle}\left( \frac{\langle n\rangle}{1+\langle
      n\rangle}\right)^{b^{\dagger}b},
\end{equation} with \begin{equation} \langle n\rangle={\rm
Tr}\{b^{\dagger}b\mu\}=\frac{A_+}{A_--A_+}\,, \end{equation} which
is the average phonon number at steady state. The coefficients
$A_+$ and $A_-$ are defined as \begin{eqnarray}\label{A}
\label{Apm} A_{\pm}=2\mbox{Re}\pg{s(\mp\nu)+{\cal D}}
\end{eqnarray} with the diffusion coefficient ${\cal
D}=\alpha\frac{\gamma}{2}\mbox{Tr}\{\sigma\da \sigma\rho_{\rm
st}\}$, and \begin{eqnarray} s(\nu)&=&\int_0^\infty d\tau e^{{\rm
i}\nu\tau} \mbox{Tr}\pg{ V_1 e^{\mathcal{L}_{I}\tau}V_1\rho_{\rm
st}} \label{Snu}. \end{eqnarray} the spectrum of fluctuation of
the dipole force~\cite{Cirac92}. Using Eq.~(\ref{V}),
Eq.~(\ref{Snu}) can be decomposed into the sum
$$s(\nu)=\varphi_L^2 s_L(\nu)+\varphi_c^2s_c(\nu)+\varphi_L
\varphi_c s_{cL}(\nu )$$ where \begin{eqnarray}\label{SL}
s_L=-\mbox{Tr}_I\{V_{1L}\pt{\mathcal{L}_{I}+{\rm i}\nu}^{-1}
V_{1L} \rho_{\rm st} \} \end{eqnarray} is the contribution of the
mechanical effect due to the laser, the term
\begin{eqnarray}\label{Sc}
s_c=-\mbox{Tr}_I\{V_{1c}\pt{\mathcal{L}_{I}+{\rm i}\nu}^{-1}
V_{1c} \rho_{\rm st} \} \end{eqnarray} the contribution of the
mechanical effect due to the resonator, and
\begin{eqnarray}\label{ScL}
s_{cL}&=&-\mbox{Tr}_I\{V_{1c}\pt{\mathcal{L}_{I}+{\rm i}\nu}^{-1}
V_{1L} \rho_{\rm st} \}\nonumber
\\
        & &-\mbox{Tr}_I\{V_{1L}\pt{\mathcal{L}_{I}+{\rm i}\nu}^{-1} V_{1c} \rho_{\rm st}
\} \end{eqnarray} is the contribution due to correlations between
the mechanical effects of laser and resonator. Depending on the
geometry of the setup and on the value of the laser and cavity
parameter, one term can be dominant over the others.

In the rest of the paper we assume a weak laser driving the atom.
This allows us to restrict the description of the coupled
dipole-cavity dynamics to the lowest {\it internal} levels, namely
the ground state $\ke{g,0_c}$, where $0_c$ indicates the vacuum
state of the cavity field, and the first two excited state
$\ke{g,1_c}$ and $\ke{e,0_c}$ with one excitation in the
dipole-cavity system. Checks of consistency show that this
approximation is reliable for the parameter regimes we considered.
Nevertheless, the center of mass motion is treated with no
approximation, and thus is a harmonic oscillator, with its
infinite ladder of equispaced levels.

\subsection{Evaluation of the spectra} \label{spectra}

In the regime of validity of the Lamb-Dicke expansion the spectra
in Eqs.~(\ref{Sat:1}) and~(\ref{Scav:1}) are evaluated by
determining the eigenvalues and eigenvectors of ${\cal L}$ in
perturbation theory in $\eta$, by solving iteratively the secular
equations~\eqref{eq:seceq} at the same order in the perturbative
expansion.

At zero order in $\eta$ internal and external degrees of freedom
are decoupled, see Eq.~(\ref{L0}). Therefore the eigenvalues of
${\cal L}_{0}$ are $\lambda_{0}=\lambda_{\rm I}+\lambda_{\rm E}$,
and the eigenelements $\varrho_{0}^{\lambda}=\rho^{\lambda_{\rm
I}}\mu^{\lambda_{\rm E}}$ where ${\cal L}_{I}\rho^{\lambda_{\rm
I}}=\lambda_{\rm I}\rho^{\lambda_{\rm I}}$ and ${\cal L}_{\rm
E}\mu^{\lambda_{\rm E}}=\lambda_{\rm E}\mu^{\lambda_{\rm E}}$. The
corresponding projector is $${\cal P}^{\lambda}_{0}=\cal
P^{\lambda_{\rm I}}\cal U^{\lambda_{\rm E}}$$ whereby their action
on the operator $X$ is defined as \begin{eqnarray*} &&{\cal
P}^{\lambda_{\rm I}}X=\rho^{\lambda_{\rm I}}{\rm
Tr}_{\rm I}\{\check{\rho}^{\lambda_{\rm I}}X\}\\
&&{\cal U}^{\lambda_{\rm E}}X=\mu^{\lambda_{\rm E}}{\rm Tr}_{\rm
E}\{\check{\mu}^{\lambda_{\rm E}}X\} \end{eqnarray*} and ${\rm
Tr}_{\rm I}$ (${\rm Tr}_{\rm E}$) denotes the trace over the
internal (external) degrees of freedom. In the limit of a
motionless atom the power spectra are characterized by the
eigenvalues of ${\cal L}_{I}$. The effect of the atomic motion
enters with the appearance of features at frequencies
$\lambda_I+\lambda_E$, with $\lambda_{\rm E}={ i}\ell\nu$ and
$\ell=0,\pm 1,\pm 2,\ldots$ At zero order in $\eta$ each
eigenspace at $\lambda_{\rm E}$ is infinitely degenerate, and the
corresponding left and right eigenelements are, for instance,
$\check{\mu}_n^{\ell}=|n+\ell\rangle\langle n|$,
$\mu^{\ell}_n=|n\rangle\langle
    n+\ell|$. These eigenelements constitute a complete and orthonormal
basis over the eigenspace at this eigenvalue. In particular, the
projector over the eigenspace at $\lambda_{\rm E}={ i}\ell\nu$ is
defined on an operator $X$ as \begin{eqnarray} \label{PE} {\cal
U}^{\lambda_{\rm E}={ i}\ell\nu}X &=&\sum_n \mu^{\ell}_n {\rm
Tr}_{\rm E}
\{\check{\mu}_n^{\ell}X\}\\
&=&\sum_n |n\rangle\langle n|X|n+\ell\rangle\langle
n+\ell|,\nonumber \end{eqnarray} where ${\rm Tr}_{\rm E}$ denotes
the trace over the external degrees of freedom.

At higher orders in the expansion in $\eta$, internal and external
degrees of freedom are coupled, and the degeneracy of the
subspaces at eigenvalue $\lambda_{\rm E}$ is
lifted~\cite{Lindberg84}. We do not report explicit results for
higher order expressions of $\lambda$, $\varrho^\lambda$ and
${\cal P}^\lambda$ here, since -- although the physical systems
are intrinsically different -- the technique for gaining these
relevant expressions is presented in detail in
Refs.~\cite{Bienert04,Bienert06}. Using the expansion in $\eta$,
the spectra in Eqs.~\eqref{Sat:1} and \eqref{Scav:1} are
decomposed as \begin{eqnarray}\label{S0S1S2} {\cal
S}\al{j}(\omega)={\cal S}_0\al{j}(\omega)+{\cal
S}_2\al{j}(\omega)+O(\eta^3). \label{eq:specexp} \end{eqnarray}
where the term ${\cal S}_\alpha\al{j}(\omega)$ is at order
$\alpha$ in the expansion in $\eta$. In zero order, the spectra
are \begin{alignat}{1} {\cal S}_0\al{\rm at}(\omega)&={\rm Re}
\sum_{\lambda_{\rm I}}\frac{1}{{\rm
i}(\omega-\omega_{L})-\lambda_{\rm I}} {\rm Tr}_{\rm
I}\left\{D_0^{\dagger} {\cal
P}^{\lambda_{\rm I}}D_0\rho_{\rm st}\right\}\label{eq:s0at}\\
{\cal S}_0\al{\rm cav}(\omega)&={\rm Re} \sum_{\lambda_{\rm
I}}\frac{1}{{\rm i}(\omega-\omega_{L})-\lambda_{\rm I}} {\rm
Tr}_{\rm I}\left\{a^{\dagger} {\cal P}^{\lambda_{\rm I}}a\rho_{\rm
st}\right\}\label{eq:s0cav} \end{alignat} and represent the
spectral signal when the atom is at rest. The first non-vanishing
correction is of second order, \begin{equation} {\cal
S}_2\al{j}(\omega)={\rm Re} \sum_{\lambda}\frac{1}{{\rm
i}(\omega-\omega_{L})-\lambda} F\al{j}(\lambda) \end{equation} and
it describes the spectral features due to mechanical effects of
the photon-atom interaction. We introduced the weighting functions
at the eigenvalue $\lambda$ \begin{alignat}{1}
F\al{\rm at}(\lambda)&=\hspace{-3mm}\sum_{\alpha+\beta+\gamma+\delta=2}\hspace{-3mm}{\rm Tr}\left\{D_\alpha^{\dagger} {\cal P}_\beta^{\lambda}D_\gamma\varrho^{\rm st}_\delta\right\}\\
F\al{\rm cav}(\lambda)&=\sum_{\alpha+\beta=2}{\rm
Tr}\left\{a^{\dagger} {\cal P}_\alpha^{\lambda}a\varrho^{\rm
st}_\beta\right\}\label{eq:Fcav} \end{alignat} where
$\alpha,\beta,\gamma,\delta$ denote the order in the Lamb-Dicke
expansion of the corresponding quantities. The resulting
expressions are quite lengthy and we abstain from reporting them
here. Their plots are presented in the next section for some
relevant parameter regimes.

\section{Results} \label{Sec:Results}

In this section we discuss the spectra of emission of atomic
fluorescence ${\cal S}\al{\rm at}$ and of cavity transmitted field
${\cal S}\al{\rm cav}$, as evaluated from Eq.~(\ref{S0S1S2}). They
are obtained when the width of the atomic transition $\gamma$ is
much larger than the trap frequency $\nu$ and $\nu\gg\kappa$, at
steady state of the cooling dynamics discussed
in~\cite{Vuletic,Zippilli05}. We remark that, in order to extract
the power spectrum, the curves presented in this section must be
multiplied with the corresponding factor $\chi\al{j}$, see
Eq.~(\ref{SS}).
\begin{figure}[!ht] \includegraphics[width=7cm]{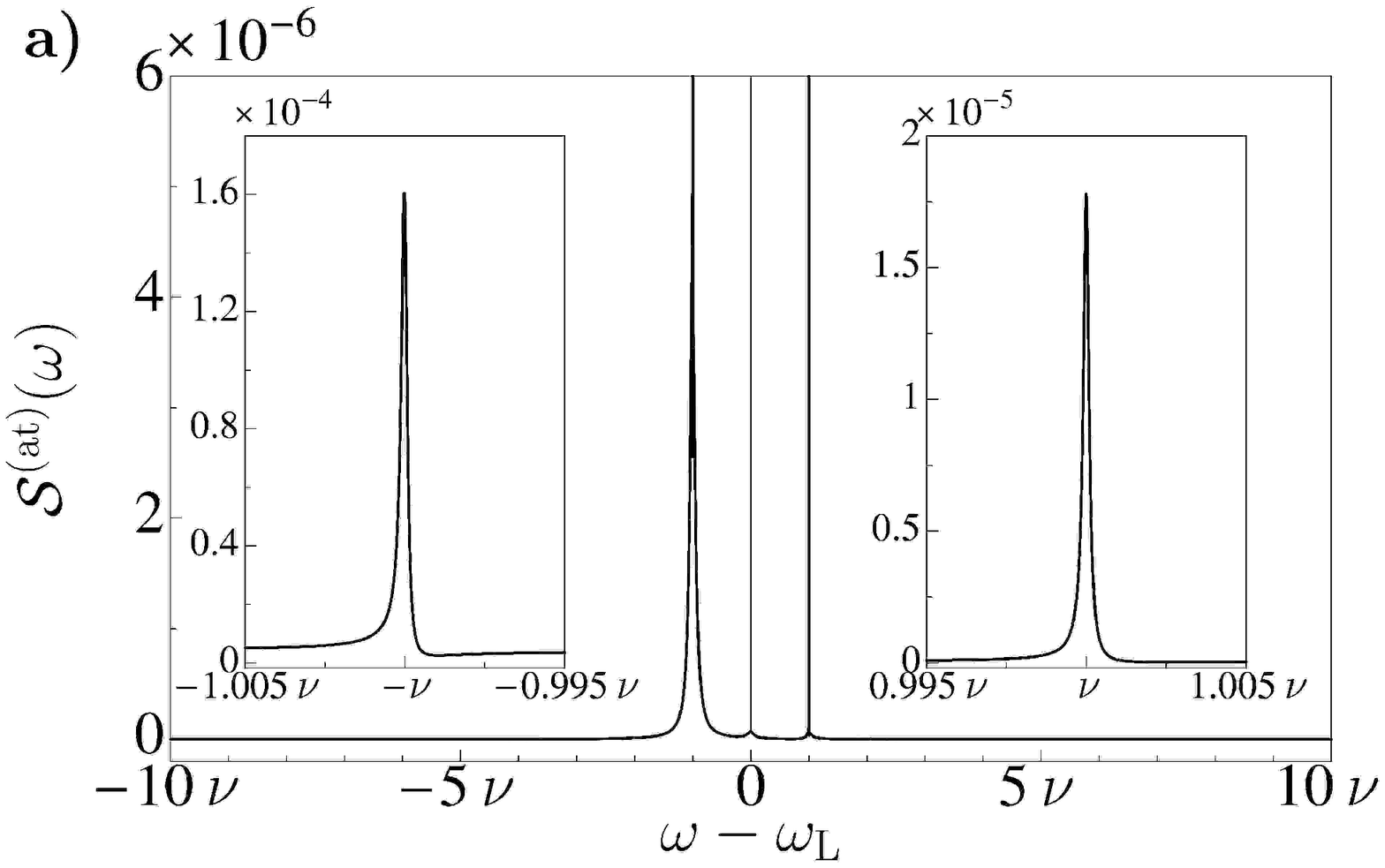}
\includegraphics[width=7cm]{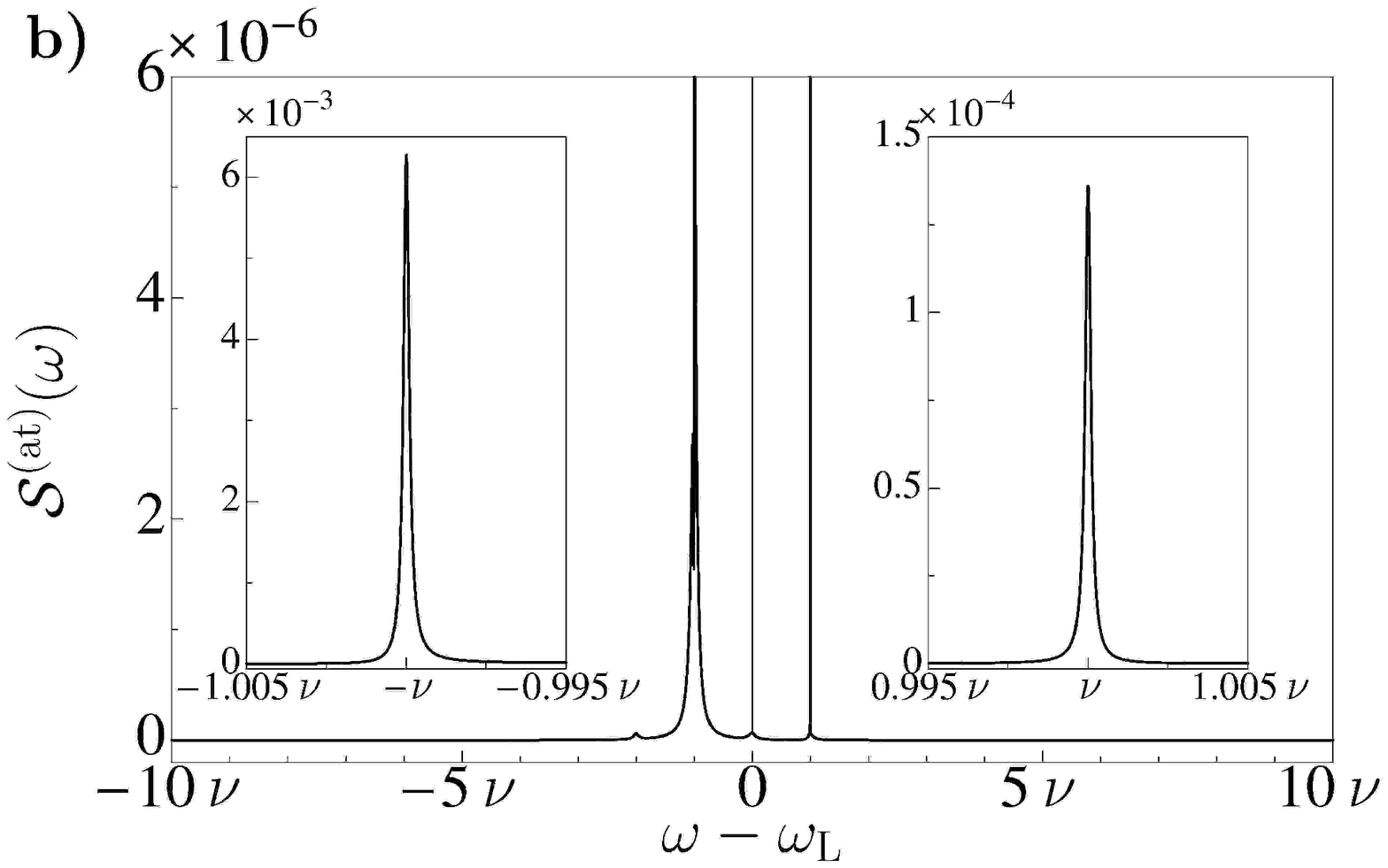}
\includegraphics[width=7cm]{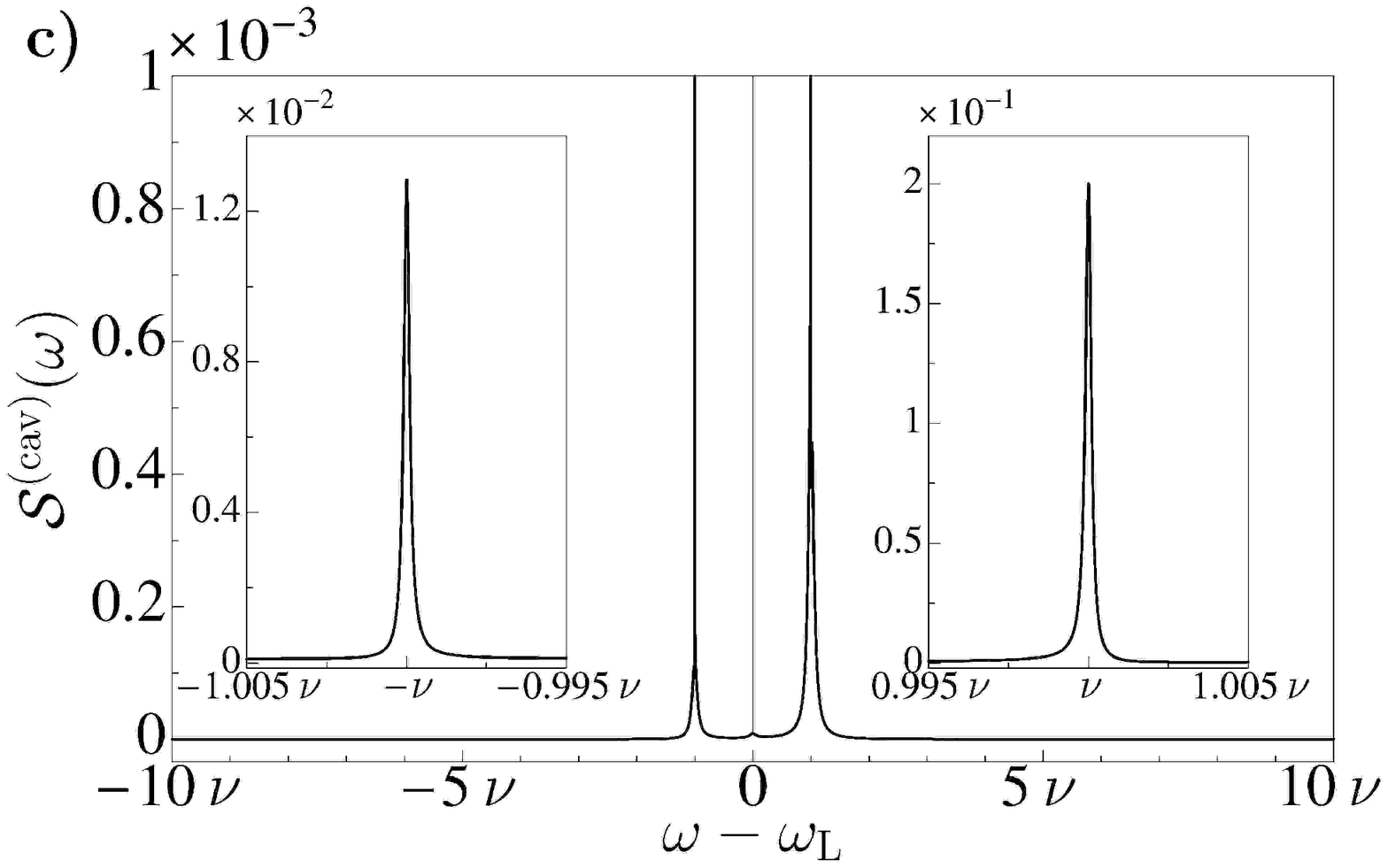} \caption{\label{fig:sbcooling}
Spectra of emission at steady state of cavity sideband cooling. a)
and b) report the spectrum of resonance fluorescence spectrum for
the detector angles $\psi=0$ and $\psi=\pi$, respectively, c) is
the spectrum at the cavity output. The plots show the
$\delta$-like elastic peak at $\omega_{\rm L}$, the narrow
sidebands of the elastic peak at $\omega_{\rm L}\pm\nu$ and small
signals of the inelastic spectrum which are due to the mechanical
coupling. The spectral components resulting from mechanical
effects depend on the detector position. The signals at
$\omega=\omega_L\pm\nu$, including the sidebands of the elastic
peak, are magnified in the insets. The parameters are
$\delta_c=-\nu$, $\Delta=500\nu$, $\Omega=5\nu$, $\tilde g=7\nu$,
$\gamma=10\nu$, $\kappa=0.1\nu$, $\eta=0.05$, $\phi=\pi/4$,
$\varphi_{\rm{c}}=\varphi_{\rm{L}}=1/\sqrt{2}$.} \end{figure}
\subsection{Spectrum for cavity assisted sideband cooling}

We first analyze the case in which the atomic motion is at steady
state of cavity sideband cooling~\cite{Vuletic,Zippilli05}. Here,
the laser is set on the red sideband transition of the cavity
mode, $\delta_c\sim-\nu$, and is far-off resonance from the atomic
dipole transition, $\abs{\Delta}\gg\gamma$. The atom-cavity states
which are mostly involved in the cooling dynamics are $\ke{g,0_c}$
and $\ke{g,1_c}$, while the excited state $\ke{e,0_c}$ is almost
empty at all stages. The corresponding spectra of emission are
displayed in Fig.~\ref{fig:sbcooling} in the frequency region
around the laser frequency. We identify features, which are
commonly observed in the resonance fluorescence of atoms,
undergoing sideband cooling in free space, namely the elastic peak
at the laser frequency and the sidebands of the elastic peak,
which are due to the harmonic motion of the atom and whose
relative height depends on the angle of emission~\cite{Cirac93}.
In fact, the atom-cavity system in this case constitutes an
effective two-level transition between the states $\ke{g,0_c}$ and
$\ke{g,1_c}$. The spectra of resonance fluorescence corresponding
to the angles $\psi=0,\pi$, where the effect of photon recoil on
the atomic motion is maximum, are plotted in
Figs.~\ref{fig:sbcooling}a) and b). Differing to the case of
sideband cooling in free space the spectrum of resonance
fluorescence exhibits an enhancement at frequency $\omega_L-\nu$,
corresponding to a heating transition, while an enhancement at the
cooling transition, at $\omega_L+\nu$, is visible in the spectrum
at the cavity output, see Fig.~\ref{fig:sbcooling}c). In fact, the
photons which cool the atom are mostly lost by cavity decay, while
correspondingly the photons which heat the motion are far off
resonance from the cavity mode, being $\kappa\ll\nu$, and are
hence mostly emitted by spontaneous decay. We checked that the
intensity at the upper and lower sidebands, integrated over the
solid angle, are equal, consistently with the fact that the system
is at steady state of the cooling dynamics. Finally, we note that
weak sidebands of the inelastic contribution at $\omega_L-\nu$ are
visible in the spectrum of resonance fluorescence, see
Fig.~\ref{fig:sbcooling}a) and b). They are located at
$\omega_{\rm L}$ and $\omega_{\rm L}-2\nu$ and are due to the
atomic motion, see~\cite{Bienert06}.

\subsection{Suppression of the inelastic spectrum}

\begin{figure}[!th] \includegraphics[width=7cm]{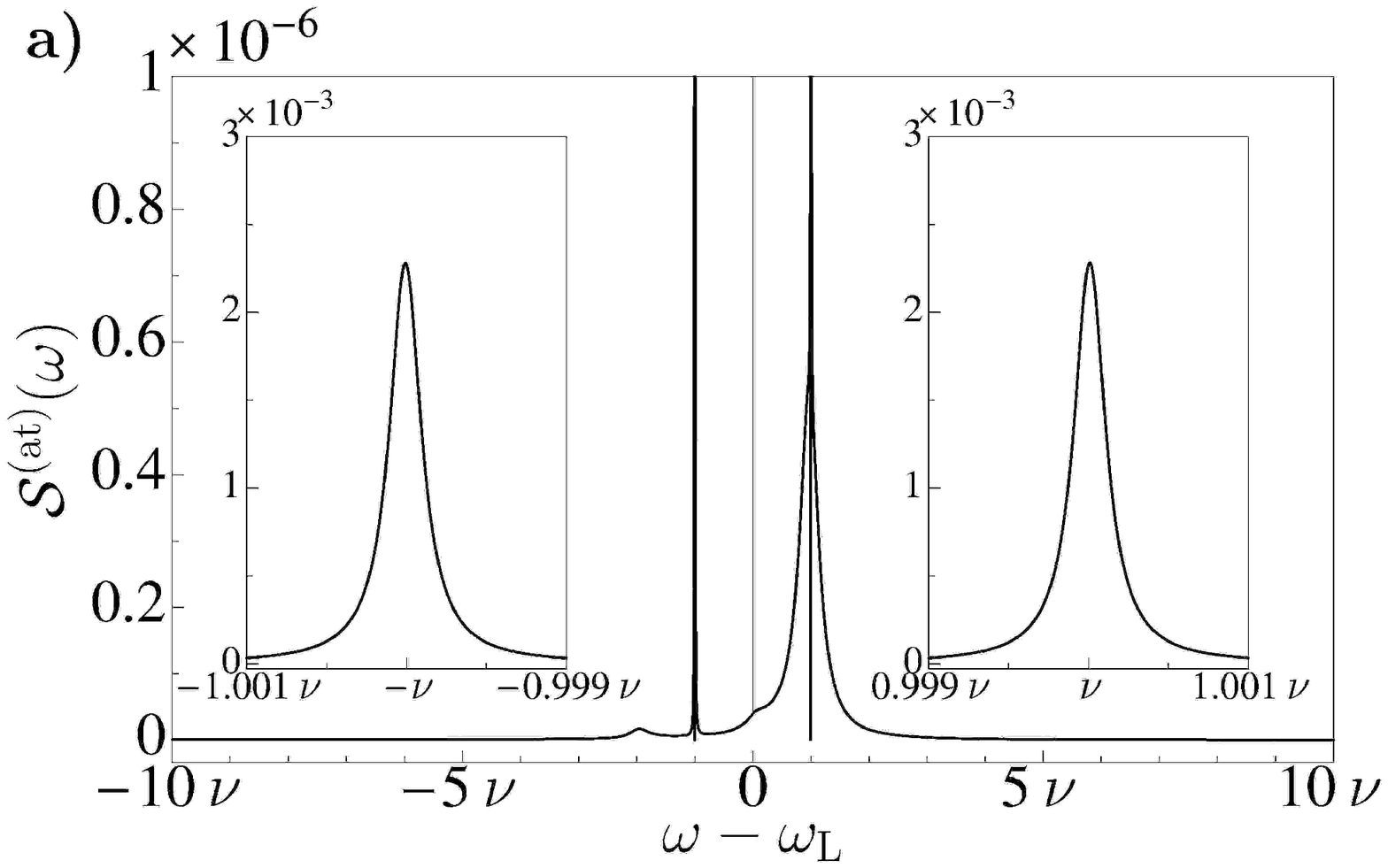}
\includegraphics[width=7cm]{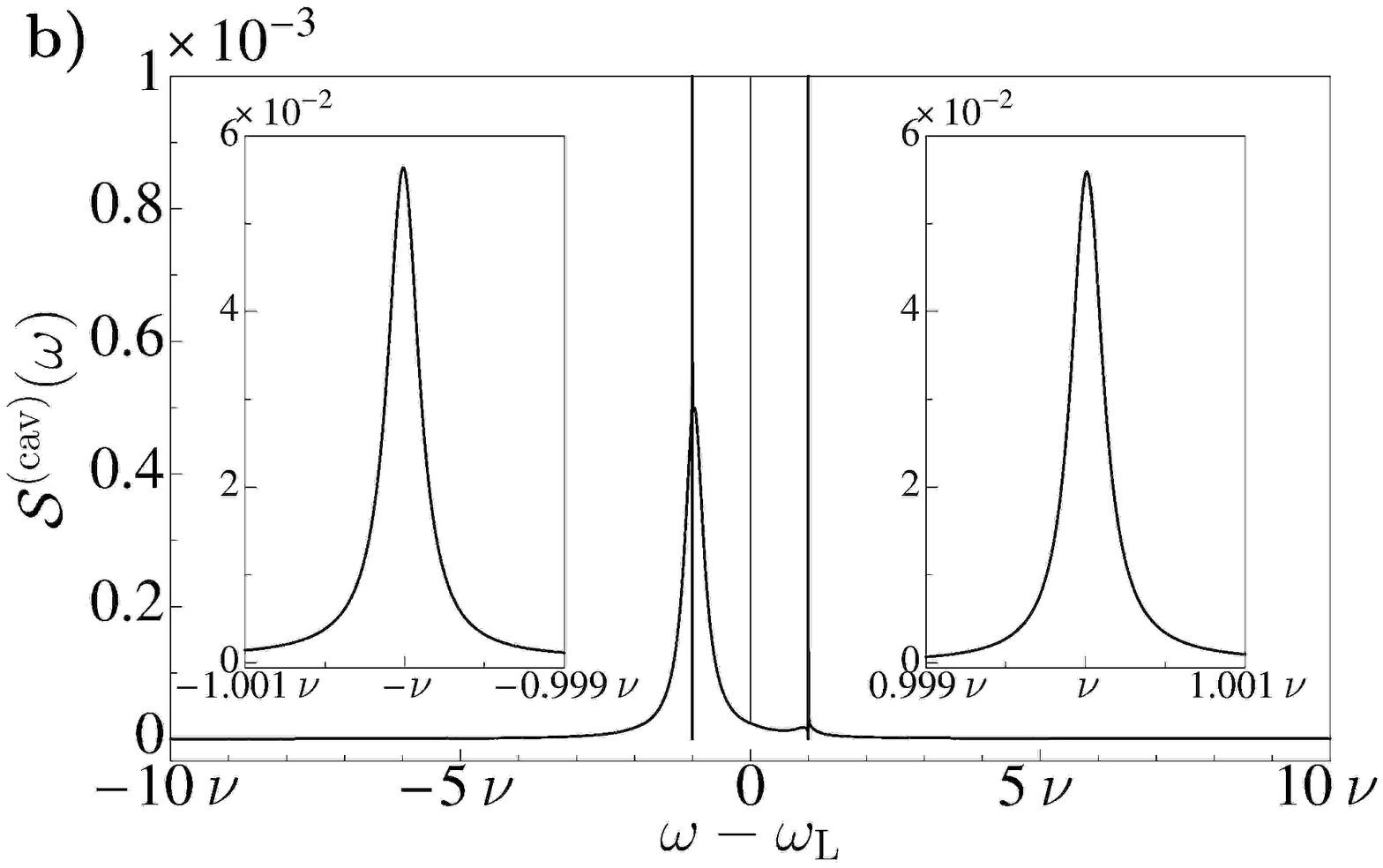} \caption{\label{fig:supofdif}
a) Spectrum of resonance fluorescence and b) spectrum of the
cavity output when the atomic motion is cooled by a scheme based
on destructive interference between laser and cavity field, which
tends to suppress atomic fluorescence~\cite{Zippilli05}. The
spectra are dominated by the motional sidebands at $\omega_{\rm
L}\pm\nu$. The broader signal in b) stems from inelastically
scattered light. In the insets the motional sidebands are
magnified. The parameters are $\delta_c=0$, $\Delta=48\nu$,
$\Omega=0.5\nu$, $\tilde g=7\nu$, $\gamma=10\nu$,
$\kappa=0.01\nu$, $\eta=0.05$, $\phi=\pi/4$,
$\varphi_{\rm{c}}=\varphi_{\rm{L}}=1/\sqrt{2}$.} \end{figure}
More striking features, which differ dramatically from the
spectrum of resonance fluorescence of a dipole in free space, are
encountered when the cooling processes are enhanced by an
interference effect between the driving laser and the field
scattered into the resonator. This phenomenon occurs when the
laser is resonant with the cavity mode, $\delta_c=0$, and for
ideal resonators ($\kappa=0$) it leads to perfect suppression of
atomic excitation for a fixed and pointlike
particle~\cite{Alsing92}. It is recovered in high finesse
resonators, provided that the cooperativity is sufficiently large.
In this case, most of the light is emitted through the cavity
mirrors~\cite{Zippilli04}.

When the finite size of the atomic wave packet is considered, the
gradient of the electromagnetic field over the atomic wave packet
gives rise to absorption, which can be shaped in order to achieve
efficient cooling~\cite{Zippilli05}. In this regime, the
excitation appearing in Fig.~\ref{Fig:3}a) is suppressed, due to
the destructive interference between the absorption paths
$|g,0_c,n\rangle\to |\pm,n\rangle$ which do not change the
motional excitation. The corresponding spectrum of resonance
fluorescence is displayed in Fig.~\ref{fig:supofdif}a). Due to the
interference effect, the elastic peak and the inelastic spectrum
contributions scale with $\kappa^2$ and $\eta^4$, respectively,
and are very small. The spectrum is dominated by the sidebands of
the elastic peak which are symmetric, and whose form is
independent of the angle of emission. In fact, due to the
destructive interference effect, the angle dependent diffusive
processes depicted in Fig.~\ref{Fig:3}a) are suppressed. This
behaviour has been also encountered in the spectrum of resonance
fluorescence of a three-level atom in $\Lambda$ configuration
driven under EIT-condition~\cite{Bienert04}. There, photon
scattering at zero order in the Lamb-Dicke expansion was
suppressed due to quantum interference in the electronic
excitations. Here, it is suppressed due to interference between
the laser and cavity field at the atomic
position~\cite{Zippilli04}.

Figure~\ref{fig:supofdif}b) displays the spectrum at the cavity
output. Here, elastic peak and inelastic spectral signals scale
with $\kappa$ and are visible. The inelastic component is centered
at the lower sideband frequency, showing that the photons which
are emitted by the resonator heat the atomic motion. The sidebands
of the elastic peak have predominant Lorentzian character and
equal height. However, a weak dispersive behaviour of the
sidebands is visible in Fig.~\ref{fig:supofdif}b), which is due to
the interference between the scattering processes depicted in
Fig.~\ref{Fig:3b}.

\subsection{Suppression of the lower sideband signal}

\begin{figure}[!htbp] \includegraphics[width=7cm]{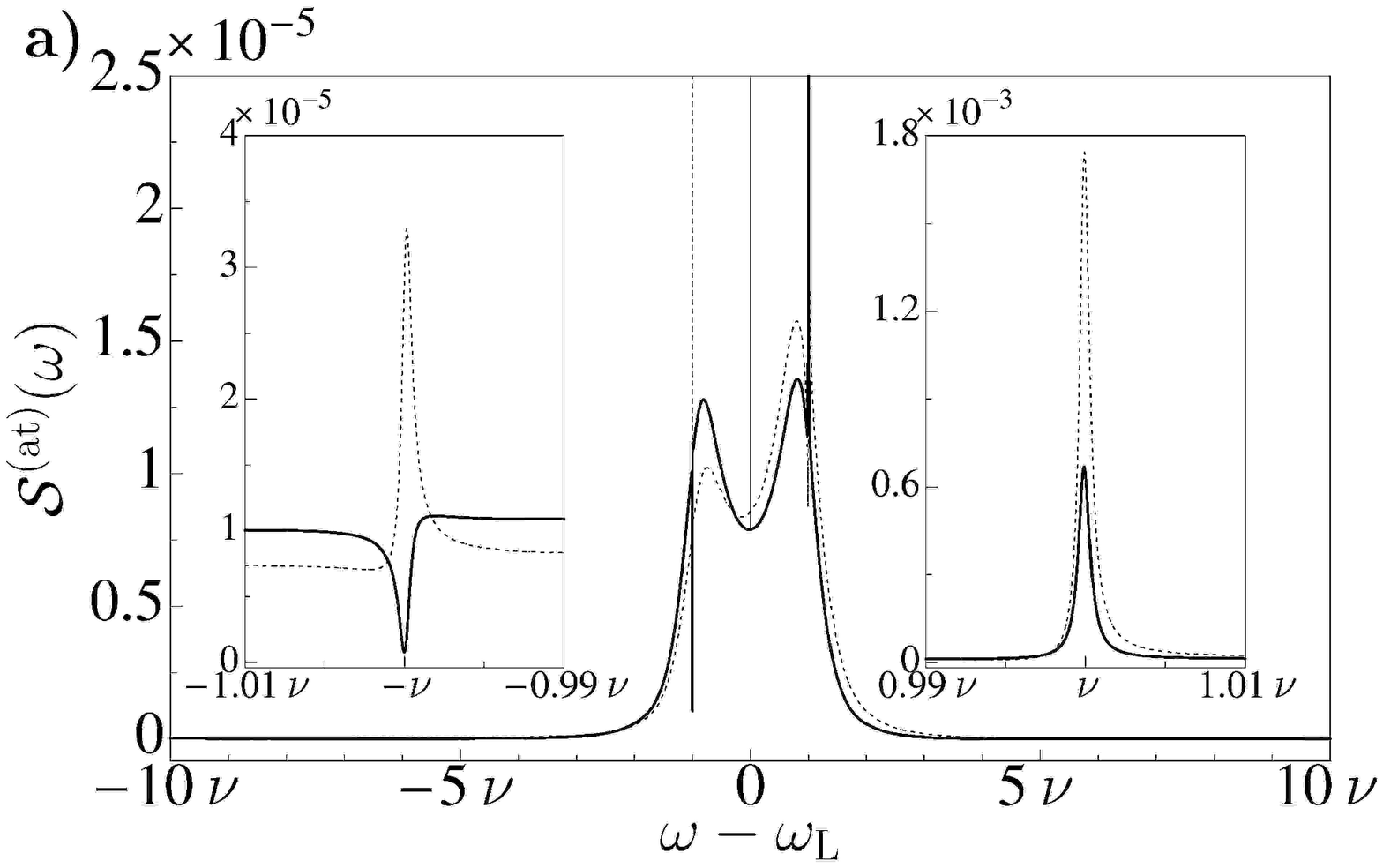}
\includegraphics[width=7cm]{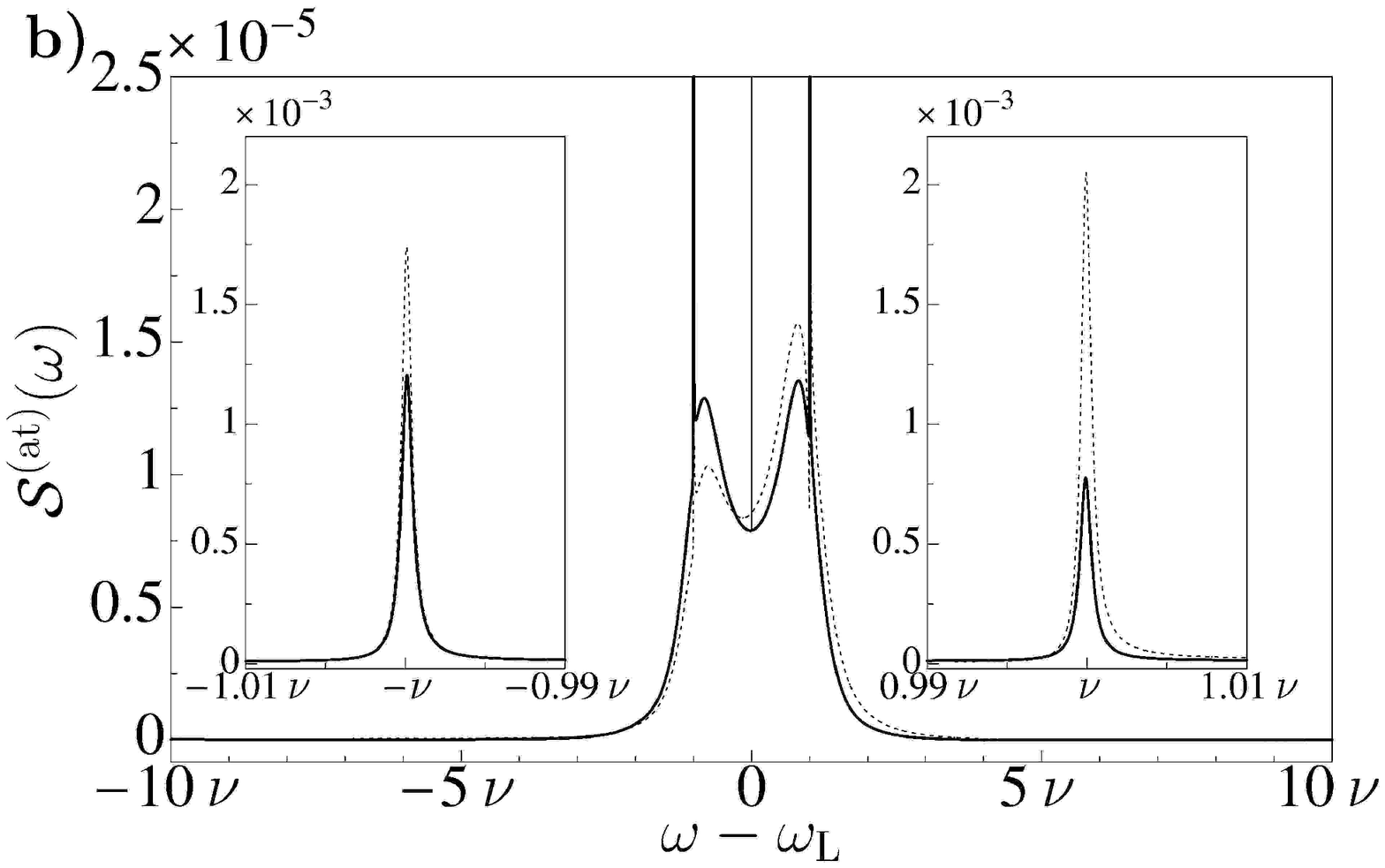}
\includegraphics[width=7cm]{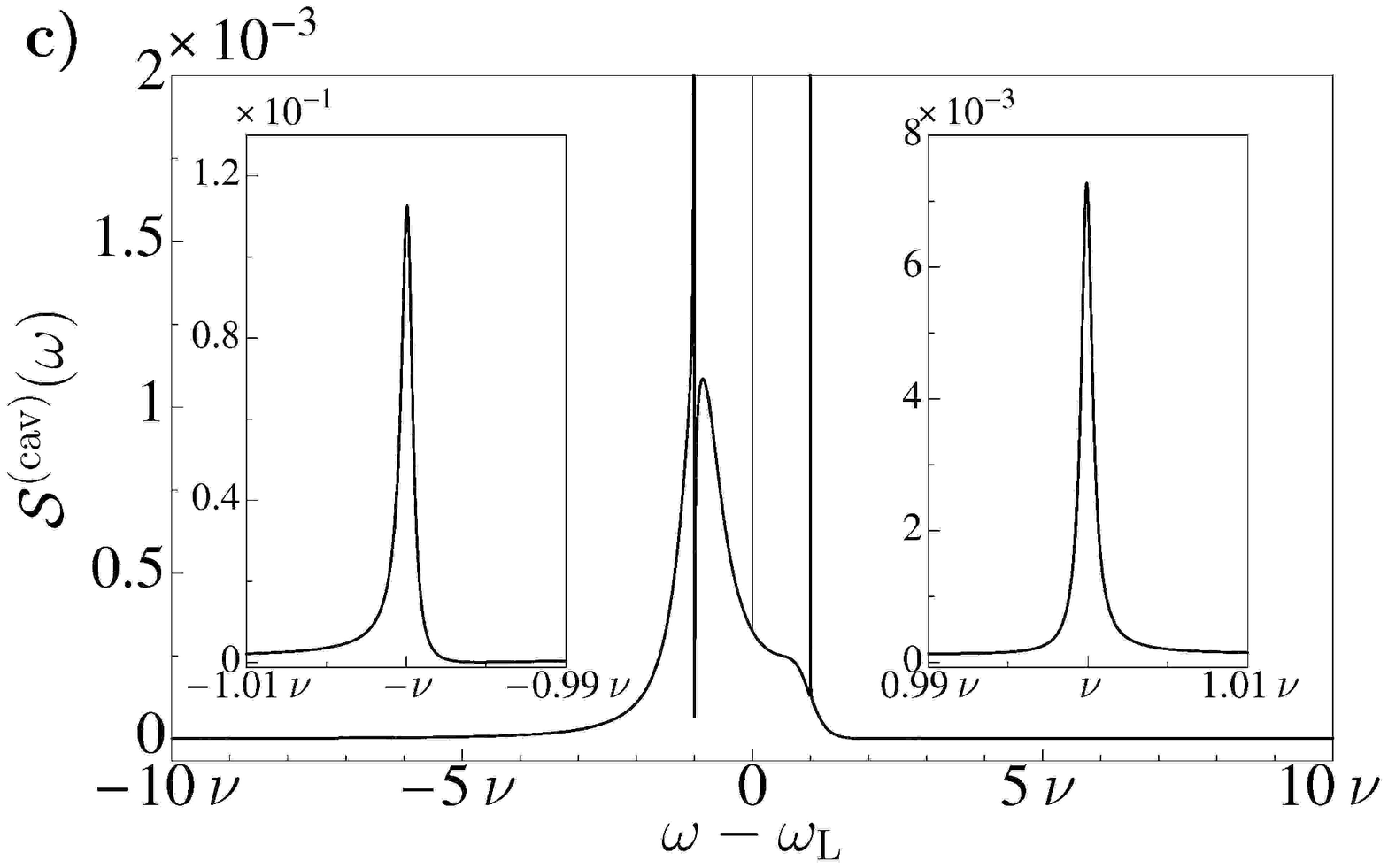} \caption{\label{fig:supheat}
Spectra of emission when the atomic motion is cooled by
suppression of the heating transition by interference between the
mechanical effects of the cavity field. The spectrum of resonance
fluorescence is displayed in a) for $\psi=\pi/2$ and b) for
$\psi=\pi$. c) Spectrum of the cavity output. The plots show the
$\delta$-like elastic peak at $\omega_{\rm L}$, the narrow
sidebands of the elastic peak at $\omega_{\rm L}\pm\nu$ and two
broad peaks resulting from inelastic processes due to the
interaction with the cavity field. The sidebands of the elastic
peak are magnified in the insets. The solid lines correspond to
$\kappa=0.01\nu$, the dotted lines to $\kappa=0.1\nu$. The
parameters are $\delta_c=\nu/2$, $\Delta=32\nu$, $\Omega=0.5\nu$,
$\tilde g=7\nu$, $\gamma=10\nu$, $\eta=0.05$, $\phi=\pi/4$,
$\varphi_{\rm{c}}=1$, $\varphi_L=0$.} \end{figure}
We now focus on the spectra obtained when cooling is based on the
suppression of the heating transition~\cite{Zippilli05}. This
behaviour stems from interference between the mechanical effects
of the cavity, originating from multiple scattering of the photon
inside the resonator, and corresponds to the destructive
interference between the scattering processes, where a photon is
spontaneously emitted and the motion is heated of one phonon. The
corresponding transitions are depicted in Fig.~\ref{Fig:3}c).

The spectra of resonance fluorescence at $\psi=\pi/2,\pi$ are
displayed in Figs.~\ref{fig:supheat}a) and~b). Here the inelastic
spectrum consists of two broad, partly overlapping peaks around
the center. These resonances correspond to transitions between the
dressed states of the cavity and the ground state at zero order in
the Lamb-Dicke expansion, see~\cite{Quang}. Other inelastic
contributions to the spectrum are very weak and lie outside the
frequency range displayed in the figures. The motional sidebands
of the elastic peak are asymmetric and depend on the angle of
emission. This can be understood in terms of interfering
scattering processes between the mechanical effects of the
spontaneously emitted photon and of the cavity field, which are
sketched in Figs.~\ref{Fig:3}c) and~a) respectively. The lower
sideband vanishes at the angle $\psi=\pi/2$, see
Fig.~\ref{fig:supheat}a), when the direction of emission is
orthogonal to the motion and there is no contribution of
spontaneous emission: In this case, the quantum interference
effect suppressing the heating transitions is clearly evident in
the spectrum, leading to the disappearance of the signal at
$\omega_L-\nu$. The dashed line corresponds to the signal when the
resonator is unstable, with $\kappa=0.1\nu$. In this case the
suppression of the heating sideband is not perfect, still this
signal is orders of magnitude smaller than the cooling sideband at
$\omega_{\rm L}+\nu$. Figure~\ref{fig:supheat}b) shows the
spectrum of resonance fluorescence when the recoil of the emitted
photon is maximum ($\psi=\pi$), and thus this process interferes
with the heating photon, giving a dispersive-like signal at
$\omega_L-\nu$.

The corresponding spectrum at the cavity output is displayed in
Fig.~\ref{fig:supheat}c) for $\kappa=0.01\nu$. The left motional
sideband at $\omega_{\rm L}-\nu$ shows a dispersive-like behavior.
The total intensity at this frequency is larger than the upper
sideband signal. This is understood, as the interference process
leading to suppression of the heating transition is taking place
between the scattering amplitudes leading to spontaneous emission.
Correspondingly, the heating photons are mostly emitted by cavity
decay and give rise to this signal.

\subsection{A monochromatic spectrum of resonance fluorescence}

\begin{figure}[!t] \includegraphics[width=7cm]{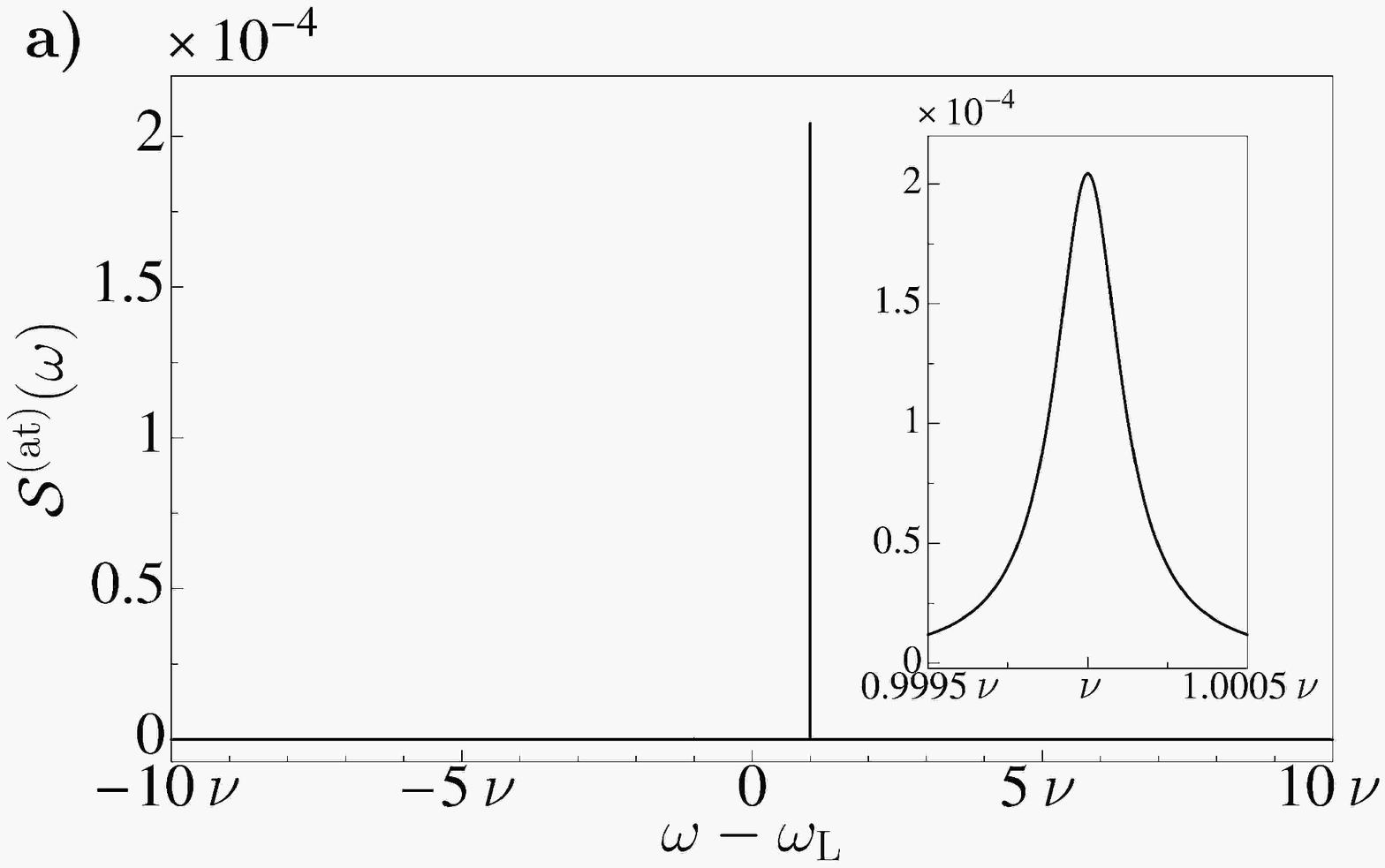}
\includegraphics[width=7cm]{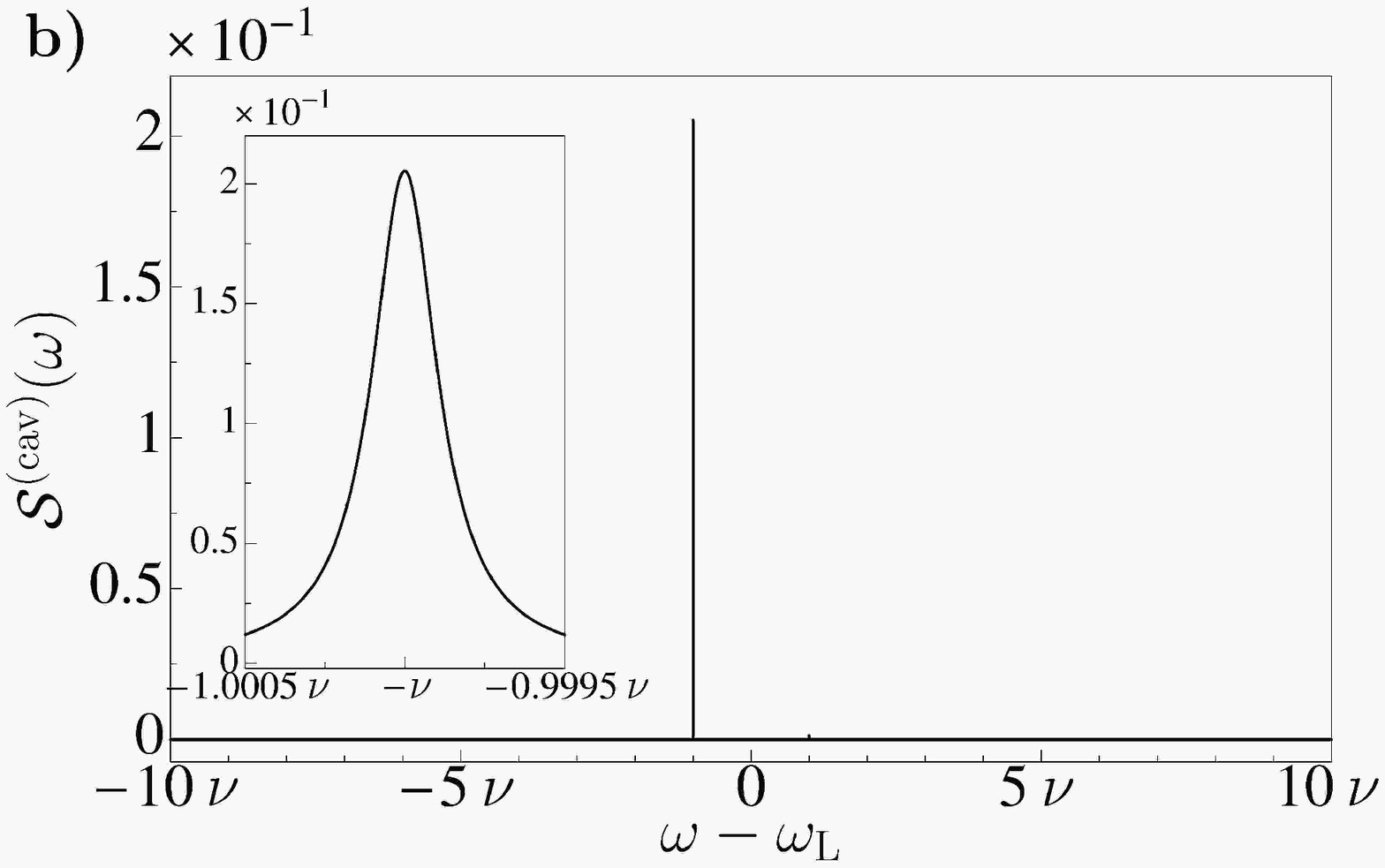}
\caption{\label{fig:supofheatdif} Spectrum of resonance
fluorescence a) and spectrum at the cavity output b) when cooling
is achieved by suppression of diffusion (by placing the atom in
the node of the laser standing wave) and heating transitions (due
to destructive interference in the mechanical effects of the
resonator).  The only visible peaks are the sidebands which are
magnified in the inset. The parameters are $\delta_c=\nu/2$,
$\Delta=23\nu$, $\Omega=0.5\nu$, $\tilde g=7\nu$, $\gamma=10\nu$,
$\kappa=0.01\nu$, $\eta=0.05$, $\varphi_{\rm{c}}=1$.} \end{figure}

Finally, we consider the spectra of emission when the atom is
driven by a standing-wave laser with the trap center placed at one
of its nodes, and the cavity parameters are such that the heating
transitions are suppressed by destructive interference. In this
configuration, the contributions to the spectra at zero order in
the Lamb Dicke expansion  defined in Eqs.~(\ref{eq:s0at})
and~(\ref{eq:s0cav}) are suppressed because of the vanishing field
at the node of the standing wave, and the resulting spectra of
scattered light are only due to mechanical effects of the
photon-atom interaction. In Figs.~\ref{fig:supofheatdif}a) and~b)
one can observe that the only relevant signals are the sidebands
of the elastic peak, which have Lorentzian form. In particular,
due to the interference effect suppressing the heating transition,
the spectrum of resonance fluorescence is only constituted by the
cooling sideband, while the spectrum at the cavity output is only
constituted by the heating motional sideband. Correspondingly, the
photons emitted by a heating transition are only detected at the
cavity output.

\section{Conclusion}\label{Sec:Conclusions}

We have analyzed the spectrum of resonance fluorescence and the
spectrum at the cavity output of a cavity cooled atom. In general,
the two spectra contain complementary information about the
scattering processes determining the atomic dynamics. In certain
parameter regimes features may appear only in one spectrum as the
result of quantum interference between transitions induced by
laser and cavity field.

These results provide further insight into the coupled dynamics of
atoms and resonators, and can be used for implementing quantum
feedback schemes, in the spirit of the proposals of
Ref.~\cite{Mancini,Steck,Zoller} and of the experimental
realization in~\cite{Bushev06}.

\acknowledgments One of the authors (M.B.) thanks the group
d'Optica at the Universitat Autonoma de Barcelona for hospitality
during completion of this work. Support by the European Commission
(EMALI, MRTN-CT-2006-035369; SCALA, Contract No.\ 015714), by the
Spanish Ministerio de Educaci\'on y Ciencia (Consolider Ingenio
2010 QOIT, CSD2006-00019; QLIQS, FIS2005-08257; Ramon-y-Cajal and
Juan de la Cierva individual fellowships), and by the
Alexander-von-Humboldt foundation, are acknowledged.

 \end{document}